\documentclass[epj]{svjour}
\usepackage{graphics}

\begin{document}

\title{Wang-Landau study of the triangular Blume-Capel ferromagnet}

\author{N.G. Fytas
\institute{Department of Materials Science, University of Patras,
26504 Patras, Greece}}

\date{Received: date / Revised version: date}

\abstract{We report on numerical simulations of the
two-dimensional Blume-Capel ferromagnet embedded in the triangular
lattice. The model is studied in both its first- and second-order
phase transition regime for several values of the crystal field
via a sophisticated two-stage numerical strategy using the
Wang-Landau algorithm. Using classical finite-size scaling
techniques we estimate with high accuracy phase-transition
temperatures, thermal, and magnetic critical exponents and we give
an approximation of the phase diagram of the model.
}
\authorrunning{N.G. Fytas} \titlerunning{Wang-Landau study of the triangular Blume-Capel ferromagnet}

\maketitle

\section{Introduction}
\label{sec:1}

The Blume-Capel model consisting of a spin-one Ising Hamiltonian
with a single-ion uniaxial crystal field
anisotropy~\cite{blume,capel} is one of the most studied models in
the communities of Statistical Mechanics and Condensed Matter
Physics. This is not only because of the relative simplicity with
which approximate calculations for this model can be carried out
and tested, as well as the fundamental theoretical interest
arising from the richness of its phase diagram, but also because
versions and extensions of the model can be applied for the
description of many different physical structures, some of them
being multi-component fluids, ternary alloys, and $^{3}$He -
$^{4}$He mixtures~\cite{lawrie}. Noteworthy, latest applications
of the Blume-Capel model include analyzes of ferrimagnets, as
recently discussed in a thorough contribution by Selke and
Oitmaa~\cite{selke-10}.

The Blume-Capel model is described by the Hamiltonian
\begin{equation}
\label{eq:1}
H=-J\sum_{<ij>}s_{i}s_{j}+\Delta\sum_{i}s_{i}^{2},
\end{equation}
where the spin variables $s_{i}$ take on the values $-1, 0$, or
$+1$, $<ij>$ indicates summation over all nearest-neighbor pairs
of sites, and $J>0$ is the ferromagnetic exchange interaction
(here we set $J=1$ and $k_{B}=1$ to fix the temperature scale).
The parameter $\Delta$ is known as the crystal-field coupling that
controls the density of vacancies ($s_{i}=0$). For
$\Delta\rightarrow -\infty$ vacancies are suppressed and the model
maps onto the Ising model.

As it is well known, the model of equation~(\ref{eq:1}) has been
analyzed, besides the original mean-field
theory~\cite{blume,capel}, by a variety of approximations and
numerical approaches, mostly on the square lattice. These include
the real space renormalization group, Monte Carlo (MC) simulations
and MC renormalization-group calculations~\cite{landau},
$\epsilon$-expansion renormalization groups~\cite{stephen}, high-
and low-temperature series calculations~\cite{fox}, a
phenomenological finite-size scaling analysis using a strip
geometry~\cite{nightingale,beale}, and, finally, the most recent
numerical approaches via the Wang-Landau
algorithm~\cite{silva,hurt,malakis}. The phase diagram of the
model consists of a segment of continuous Ising-like transitions
at high temperatures and low values of the crystal field which
ends at a tricritical point, where it is joined with a second
segment of first-order transitions between ($\Delta_{t},T_{t}$)
and ($\Delta=zJ/2,T=0$), where the subscript $t$ refers to the
tricritical point and $z$ denotes the coordination number of the
considered lattice.

In the present paper we are interested in the critical properties
of the Blume-Capel model embedded in the triangular lattice
($z=6$). For this particular case, Mahan and Girvin~\cite{mahan}
were the first to apply position-space renormalization-group
methods to solve the model. These authors estimated the critical
frontier of the model with the location of the tricritical point
$(\Delta_{t},T_{t})$ at $(2.686,1.493)$. Many years later, Du et
al.~\cite{du} performed a further sophisticated analytical
calculation using an expanded Bethe-Peierls approximation to find
respectively the estimates $(\Delta_{t},T_{t})=(2.818,1.598)$.
Here, we employ MC simulations to investigate several aspects of
the phase diagram of the model and refine mean-field-type
estimates, which, as is well known, suffer from large
errors.~\footnote{Note here that the estimation of Du et al. for
the corresponding tricritical crystal field value of the square
lattice is $\Delta_{t}=1.906$, whereas the most accurate MC
estimation in the literature is $\Delta_{t}=1.96581...$ by Qian et
al.~\cite{qian}.} Our extensive simulations follow a sophisticated
numerical scheme, as described in the following Section, that
enabled us to perform a detailed finite-size scaling analysis of
the critical properties of the model in both its first- and
second-order phase transition regimes. These results, together
with an approximation of the phase diagram of the model and the
estimation of the tricritical value of $\Delta_{t}$ via a new
scheme are presented in Section~\ref{sec:3}. This contribution is
ended in Section~\ref{sec:4}, where a brief summary of our
conclusions is given.

\section{Outline of the Numerical Approach}
\label{sec:2}

In the last few years we have used an entropic sampling
implementation of the Wang-Landau algorithm~\cite{wang} to study
some simple~\cite{malakis04}, but also some more complex
systems~\cite{malakis06,fytas08a,fytas08c}. One basic ingredient
of this implementation is a suitable restriction of the energy
subspace for the implementation of the Wang-Landau algorithm. This
was originally termed as the critical minimum energy subspace
restriction~\cite{malakis04} and it can be carried out in many
alternative ways, the simplest being that of observing the
finite-size behavior of the tails of the energy probability
density function of the system~\cite{malakis04}.

Complications that may arise in complex systems, i.e. random
systems or systems showing a first-order phase transition, can be
easily accounted for by various simple modifications that take
into account possible oscillations in the energy probability
density function and expected sample-to-sample fluctuations of
individual realizations. In our recent
papers~\cite{malakis,fytas08a,fytas08c}, we have presented details
of various sophisticated routes for the identification of the
appropriate energy subspace $(E_{1},E_{2})$ for the entropic
sampling of each realization. In estimating the appropriate
subspace from a chosen pseudocritical temperature one should be
careful to account for the shift behavior of other important
pseudocritical temperatures and extend the subspace appropriately
from both low- and high-energy sides in order to achieve an
accurate estimation of all finite-size anomalies. Of course,
taking the union of the corresponding subspaces, insures accuracy
for the temperature region of all studied pseudocritical
temperatures.

The up to date version of our implementation uses a combination of
several stages of the Wang-Landau process. First, we carry out a
starting (or preliminary) multi-range (multi-R) stage, in a very
wide energy subspace. This preliminary stage is performed up to a
certain level of the Wang-Landau random walk. The Wang-Landau
refinement is $G(E)\rightarrow f\cdot G(E)$, where $G(E)$ is the
density of states (DOS) and we follow the usual modification
factor adjustment $f_{j+1}=\sqrt{f_{j}}$ and $f_{1}=e$. The
preliminary stage may consist of the levels : $j=1,\ldots,j=18$
and to improve accuracy the process may be repeated several times.
However, in repeating the preliminary process and in order to be
efficient, we use only the levels $j=13,\ldots,18$ after the first
attempt, using as starting DOS the one obtained in the first
random walk at the level $j=12$. From our experience, this
practice is almost equivalent to simulating the same number of
independent Wang-Landau random walks. Also in our recent studies
we have found out that is much more efficient and accurate to
loosen up the originally applied very strict flatness
criteria~\cite{malakis04}. Thus, a variable flatness process
starting at the first levels with a very loose flatness criteria
and assuming at the level $j=18$ the original strict flatness
criteria is nowadays used. After the above described preliminary
multi-R stage, in the wide energy subspace, one can proceed in a
safe identification of the appropriate energy subspace using one
or more alternatives outlined in reference~\cite{malakis04}.

The process continues in two further stages (two-stage process),
using now mainly high iteration levels, where the modification
factor is very close to unity and there is not any significant
violation of the detailed balance condition during the Wang-Landau
process. These two stages are suitable for the accumulation of
histogram data (for instance energy-magnetization histograms),
which can be used for an accurate entropic calculation of
non-thermal thermodynamic parameters, such as the order parameter
and its susceptibility~\cite{malakis04}. In the first (high-level)
stage, we follow again a repeated several times (typically $\sim
5-10$) multi-R Wang-Landau approach, carried out now only in the
restricted energy subspace. The Wang-Landau levels may be now
chosen as $j=18,19,20$ and as an appropriate starting DOS for the
corresponding starting level the average DOS of the preliminary
stage at the starting level may be used. Finally, the second
(high-level) stage is applied in the refinement Wang-Landau levels
$j=j_{i},\ldots,j_{i}+3$ (typically $j_{i}=21$), where we usually
test both an one-range (one-R) or a multi-R approach with large
energy intervals. In the case of the one-R approach we have found
very convenient and in most cases more accurate to follow the
Belardinelli and Pereyra~\cite{belardinelli07} adjustment of the
Wang-Landau modification factor according to the rule $\ln f\sim
t^{-1}$. Finally, it should be also noted that by applying in our
scheme a separate accumulation of histogram data in the starting
multi-R stage (in the wide energy subspace) offers the opportunity
to inspect the behavior of all basic thermodynamic functions in an
also wide temperature range and not only in the neighborhood of
the finite-size anomalies. The approximation outside the dominant
energy subspace is not of the same accuracy with that of the
restricted dominant energy subspace but is good enough for the
observation of the general behavior and provides also a route of
inspecting the degree of approximation.

In the present work, the above described numerical approach was
used to estimate the properties of the triangular Blume-Capel
model for lattice sizes in the range $L\in \{20-100\}$ for all
values of the crystal field considered. For each pair $(L,
\Delta)$, $50$ independent runs were performed. We close this
outline of our numerical scheme with some comments concerning
statistical errors. Even for the larger lattice size studied here
($L=100$), and depending on the thermodynamic parameter, the
statistical errors of the Wang-Landau method were found to be of
reasonable magnitude and in some cases to be of the order of the
symbol sizes, or even smaller. Thus, the error bars shown in all
our figures in the following Section have been estimated as
standard deviations from the ensemble of the $50$ independent
runs.

\section{Results and Analysis}
\label{sec:3}

\begin{figure}
\resizebox{1 \columnwidth}{!}{\includegraphics{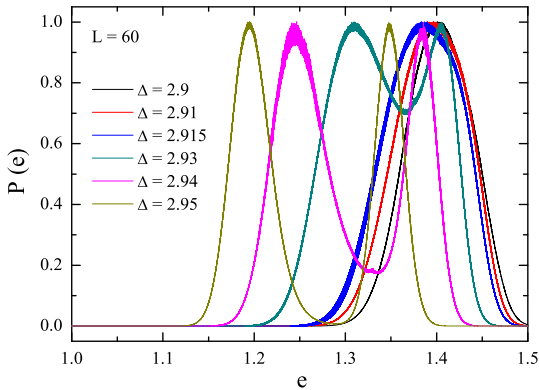}}
\caption{Energy probability density functions for a lattice size
$L=60$ and several values of the crystal field, as indicated. For
the cases of a double-peaked structure we have used the
temperature where the two peaks are of equal height, whereas in
the other single-peaked cases, we plot the densities at the
corresponding pseudocritical temperatures of the specific heat.}
\label{fig:1}
\end{figure}

\begin{figure}
\resizebox{1 \columnwidth}{!}{\includegraphics{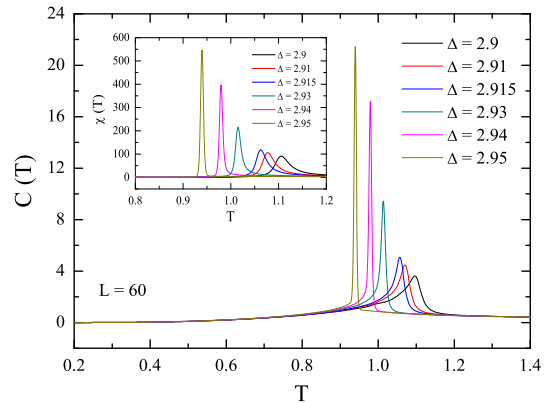}}
\caption{Corresponding specific heats (main panel) and magnetic
susceptibilities (inset) as a function of temperature. This figure
is complementary to figure~\ref{fig:1}.} \label{fig:2}
\end{figure}

We present in this Section our numerical results and analysis for
the triangular Blume-Capel model. For convenience we separate our
discussion in three parts. The first (subsection~\ref{sec:3a})
refers to the second-order phase transition regime of the model,
the following (subsection~\ref{sec:3b}) to the first-order phase
transition regime, and finally in the last subsection~\ref{sec:3c}
we give our approximation on the phase diagram of the model
together with a novel estimation approach of the tricritical
crystal field value $\Delta_{t}$. Preliminary runs, as shown in
figures~\ref{fig:1} and \ref{fig:2}, indicate that the location of
the tricritical point of the model is in the regime $\Delta >
2.915$. More specifically, in figure~\ref{fig:1} we plot the
energy probability density function $P(e)$, where $e=H/L^{d}$ and
$d$ the lattice dimensionality ($d=2$ in the present study), for a
lattice size $L=60$ for several values of the crystal field in the
regime $\Delta=2.9-2.95$. The double-peaked structure in the
energy probability density function, signaling a first-order phase
transition~\cite{binder84,binder87}, appears after the value
$\Delta=2.915$. Respectively, figure~\ref{fig:2} illustrates the
corresponding specific heats $C$ (main panel) and magnetic
susceptibilities $\chi$ (inset) as a function of the temperature
for the same values of the crystal field of figure~\ref{fig:1}.
Again the sharp peak in both quantities, characteristic of a
first-order phase transition, is observed in the same
$\Delta$-regime.

Thus, using the above information, we have chosen to simulate the
values $\Delta\in \{0,0.5,1,1.5,2,2.5,2.75\}$ in the second-order
phase transition regime of the model and the value $\Delta=2.95$
in the first-order phase transition regime, respectively.

\subsection{Second-order phase transition regime}
\label{sec:3a}

\begin{figure}
\resizebox{1 \columnwidth}{!}{\includegraphics{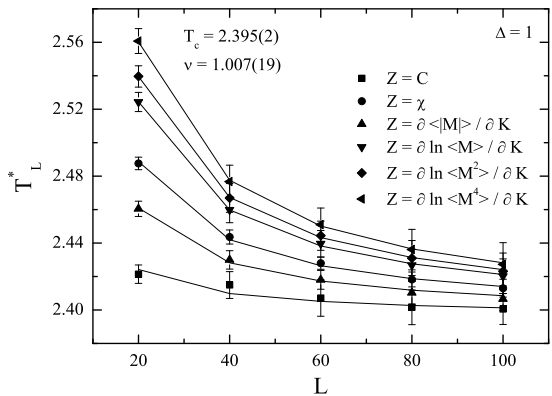}}
\caption{Simultaneous fitting of the form~(\ref{eq:4}) of the six
pseudocritical temperatures defined in the text for the case
$\Delta=1$.} \label{fig:3}
\end{figure}

The triangular Blume-Capel model at the crystal field value
$\Delta < \Delta_{t}$, undergoes a second-order transition between
the ferromagnetic and paramagnetic phases, expected to be in the
universality class of the simple $d=2$ Ising model. In the
following, we present the finite-size scaling analysis of our
numerical data for this case, to verify this expectation. In
figure~\ref{fig:3} we give an example, for the case $\Delta=1$, of
the shift behavior of the pseudocritical temperatures
corresponding to the peaks of the following six quantities:
specific heat $C$, magnetic susceptibility $\chi$, derivative of
the absolute order parameter with respect to the inverse
temperature $K=1/T$~\cite{ferrenberg91}
\begin{equation}
\label{eq:2} \frac{\partial \langle |M|\rangle}{\partial
K}=\langle |M|H\rangle-\langle |M|\rangle\langle H\rangle,
\end{equation}
and logarithmic derivatives of the first-, second-, and
fourth-order powers of the order parameter with respect to the
inverse temperature $K=1/T$~\cite{ferrenberg91}
\begin{equation}
\label{eq:3} \frac{\partial \ln \langle M^{n}\rangle}{\partial
K}=\frac{\langle M^{n}H\rangle}{\langle M^{n}\rangle}-\langle
H\rangle.
\end{equation}
Fitting our data to the expected power-law behavior
\begin{equation}
\label{eq:4} T_{L}^{\ast}=T_{c}+bL^{-1/\nu},
\end{equation}
we find the critical temperature to be $T_{c}=2.395(2)$ and the
correlation length exponent $\nu=1.007(19)$, in agreement with the
value of $\nu=1$ of the simple Ising model. The analysis presented
in figure~\ref{fig:3} has been performed for all the values of the
crystal field considered in the present paper to estimate
transition temperatures, but are not shown here for brevity. These
values will be used in the sequel in the construction of the phase
diagram of the model.

\begin{figure}
\resizebox{1 \columnwidth}{!}{\includegraphics{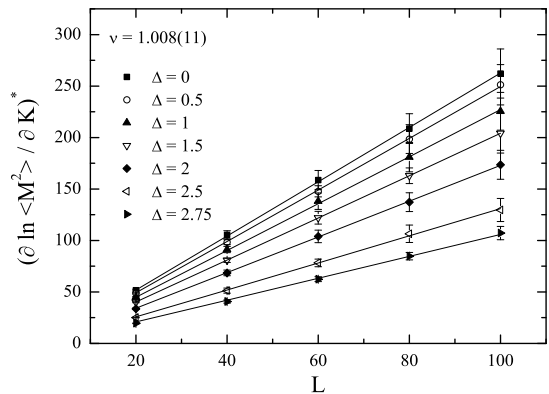}}
\caption{Simultaneous fitting of the maxima of the second-order
logarithmic derivative of the order parameter for $\Delta\in
[0-2.75]$.} \label{fig:4}
\end{figure}

Figure~\ref{fig:4} gives an alternative estimation of the
correlation length exponent via the finite-size scaling behavior
of the logarithmic derivatives of the order parameter, whose
maxima are expected to scale (in a second-order phase transition)
as $\sim L^{1/\nu}$ with the system size~\cite{ferrenberg91}. We
chose to show in figure~\ref{fig:4} the case $n=2$ of
equation~(\ref{eq:3}). A simultaneous fitting for all the values
of the crystal field considered gives an excellent estimate of
$1.008(11)$ for the critical exponent $\nu$ of the correlation
length.

\begin{figure}
\resizebox{1 \columnwidth}{!}{\includegraphics{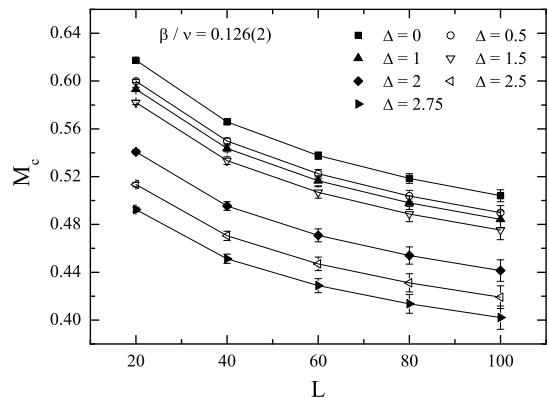}}
\caption{Simultaneous fitting of the critical order-parameter data
for $\Delta\in [0-2.75]$.} \label{fig:5}
\end{figure}

\begin{figure}
\resizebox{1 \columnwidth}{!}{\includegraphics{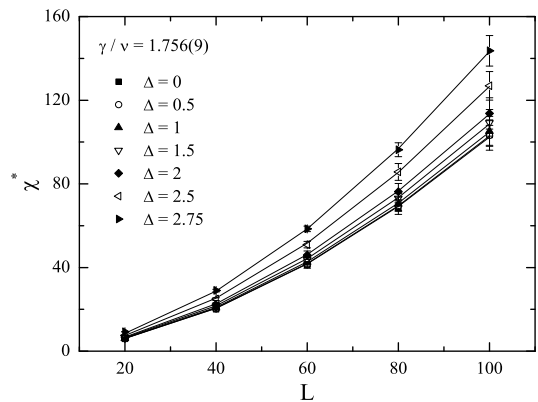}}
\caption{Simultaneous fitting of the maxima of the magnetic
susceptibility for $\Delta\in [0-2.75]$.} \label{fig:6}
\end{figure}

We proceed with figures~\ref{fig:5} and \ref{fig:6} to estimate
the magnetic exponent ratios of the triangular Blume-Capel model.
In particular in figure~\ref{fig:5} we present the order-parameter
data at the estimated critical temperatures as a function of the
lattice size. The lines show a simultaneous fitting for the whole
spectrum of $\Delta$-values of the form $M_{c} \sim
L^{-\beta/\nu}$ that gives an estimate $0.126(2)$ for the critical
exponent ratio $\beta/\nu$, in very good agreement with expected
value $0.125$ of the pure Ising case. Respectively, in
figure~\ref{fig:6} we plot the maxima of the magnetic
susceptibility, again as a function of the lattice size. As in
figure~\ref{fig:5}, the lines show a simultaneous power-law
fitting of the form $\chi^{\ast} \sim L^{\gamma/\nu}$ that gives
an estimate $1.756(9)$ for the magnetic exponent ratio
$\gamma/\nu$, also in very good agreement with expected value
$1.75$ of the pure Ising case.

\begin{figure}
\resizebox{1 \columnwidth}{!}{\includegraphics{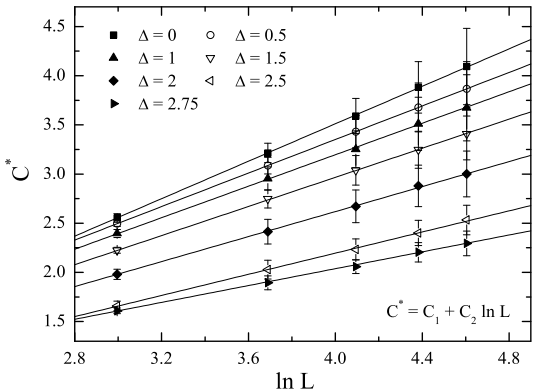}}
\caption{Finite-size scaling of the specific heat maxima. The
lines show a simultaneous fitting of the form~(\ref{eq:5}) for
$\Delta\in [0-2.75]$. Note the logarithmic scale in the horizontal
axis.} \label{fig:7}
\end{figure}

Closing this subsection, we deal with the most intriguing
thermodynamic quantity in the study of spin models: the specific
heat. The specific heat is an extremely sensitive quantity and it
is well known that it is, at least in some cases, a very hard task
to identify with good accuracy its scaling behavior. However, in
the present work, our careful numerical implementation of the
Wang-Landau scheme and the repeated sampling gave very well
numerical data, as can be shown in figure~\ref{fig:7}, where we
present the finite-size scaling behavior of the specific heat
maxima. The solid lines show a excellent simultaneous logarithmic
fitting of the form
\begin{equation}
\label{eq:5} C^{\ast}=C_{1}+C_{2} \ln L.
\end{equation}
Note here that, the expected logarithmic divergence of the
specific heat is very well obtained even from the smaller lattice
sizes shown.

Summarizing, the results and analysis presented in this subsection
for the triangular Blume-Capel model at its second-order phase
transition regime are in full agreement with universality
arguments that place the Blume-Capel model for $\Delta <
\Delta_{t}$ in the Ising universality class.

\subsection{First-order phase transition regime}
\label{sec:3b}

We now consider the case $\Delta=2.95$, for which the model
undergoes a first-order transition between the ferromagnetic and
paramagnetic phases. Our first attempt to elucidate the
first-order transition features of the present model will closely
follow previous analogous studies carried out on the $q \geq 5$
Potts model~\cite{challa86,lee90,borgs92,janke93} and also our
studies of the corresponding square lattice Blume-Capel
model~\cite{malakis} and the triangular Ising model with nearest-
and next-nearest-neighbor antiferromagnetic
interactions~\cite{malakis07}. As it is well known from the
existing theories of first-order transitions, all finite-size
contributions enter in the scaling equations in powers of the
system size $L^{d}$~\cite{fisher82}. This holds for the general
shift behavior and also for the finite-size scaling behavior of
the peaks of various energy cumulants and of the magnetic
susceptibility. It is also well known, as mentioned above, that
the double-peaked structure of the energy probability density
function $P(e)$ is signaling the emergence of the expected two
delta-peak behavior in the thermodynamic limit, for a genuine
first-order phase transition~\cite{binder84,binder87}, and with
increasing lattice size the barrier between the two peaks should
steadily increase. According to the arguments of Lee and
Kosterlitz~\cite{lee90} the quantity $\Delta
F(L)/L^{d-1}=[k_{B}T\ln{(P_{max}/P_{min})}]/L$, where $P_{max}$
and $P_{min}$ are the maximum and minimum energy probability
density function values at the temperature $T_{h}$ where the two
peaks are of equal height, should tend to a non-zero value.

\begin{figure}
\resizebox{1 \columnwidth}{!}{\includegraphics{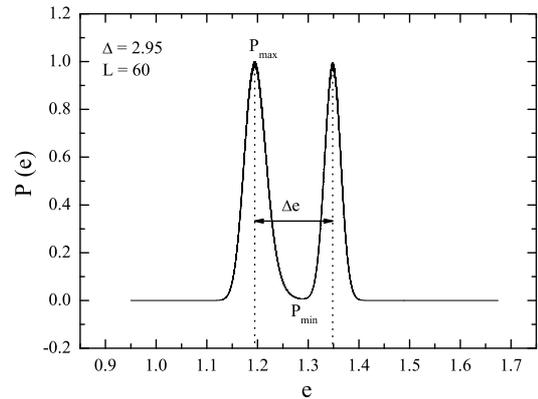}}
\caption{The double-peaked structure of the energy probability
density function at $T=T_{h}$ for $\Delta=2.95$ and $L=60$.}
\label{fig:8}
\end{figure}

\begin{figure}
\resizebox{1 \columnwidth}{!}{\includegraphics{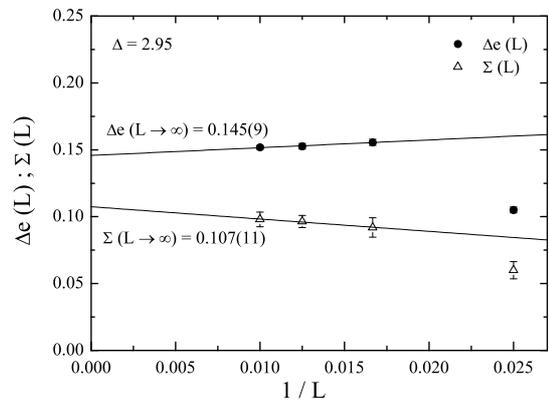}}
\caption{Limiting behavior for $\Delta=2.95$ of the surface
tension $\Sigma (L)$, defined in the text, and the latent heat
$\Delta e(L)$, shown in figure~\ref{fig:8}. The solid lines are
linear fittings to $L\rightarrow \infty$.} \label{fig:9}
\end{figure}

Figure~\ref{fig:8} shows the pronounced double-peaked structure of
the energy probability density function of the model at the
temperature $T=T_{h}$ where the two peaks are of equal height for
$L=60$. From the double-peaked energy probability density function
one can estimate the surface tension $\Sigma(L)=\Delta F(L)/L$ and
the latent heat $\Delta e(L)$, whose values remain finite for a
genuine first-order transition. Figure~\ref{fig:9} shows the
limiting behavior of these two quantities and verifies the
persistence of the first-order character of the transition at
$\Delta=2.95$. The limiting values of $\Sigma(L)$ and $\Delta
e(L)$ are given in the graph by extrapolating at the larger
lattice sizes studied.

\begin{figure}
\resizebox{1 \columnwidth}{!}{\includegraphics{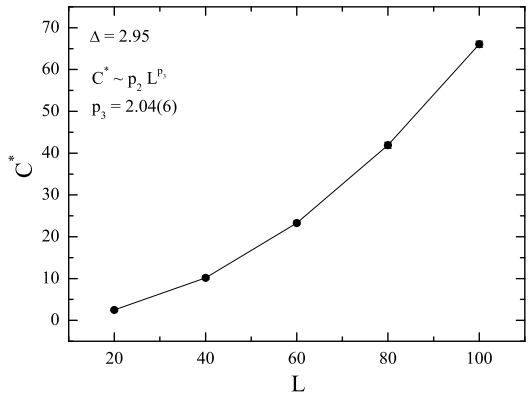}}
\caption{Finite-size scaling behavior of the specific heat peaks
at $\Delta=2.95$ giving a clear $L^{d}$ divergence.}
\label{fig:10}
\end{figure}

\begin{figure}
\resizebox{1 \columnwidth}{!}{\includegraphics{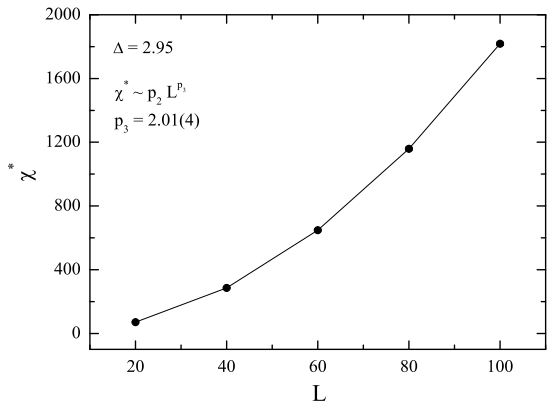}}
\caption{Finite-size scaling behavior of the susceptibility maxima
at $\Delta=2.95$. A clear $L^{d}$ divergence is observed as in the
case of the specific heat.} \label{fig:11}
\end{figure}

\begin{figure}
\resizebox{1 \columnwidth}{!}{\includegraphics{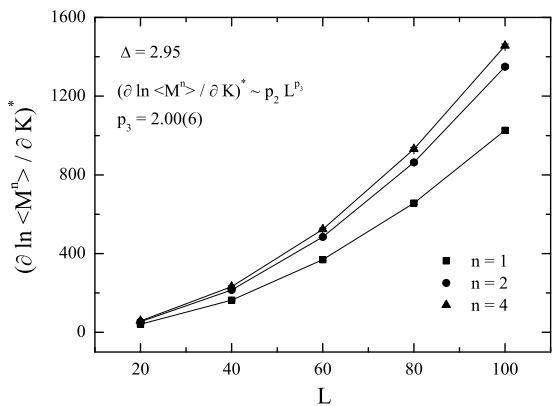}}
\caption{Simultaneous fitting of the maxima of the three
logarithmic derivatives of the order parameter~(\ref{eq:3}) for
$n=1$, $2$, and $4$, at $\Delta=2.95$. As in figures~\ref{fig:10}
and \ref{fig:11} a clear $L^{d}$ divergence is observed.}
\label{fig:12}
\end{figure}

Figures~\ref{fig:10} and \ref{fig:11} illustrate that the
traditionally used divergences in finite-size scaling of the
specific heat $C$ and susceptibility $\chi$ follow very well a
power-law behavior of the form $\sim L^{d}$, as expected for
first-order transitions~\cite{binder84,binder87}. Furthermore,
figure~\ref{fig:12} demonstrates that the divergences
corresponding to the first-, second-, and fourth-order logarithmic
derivatives of the order parameter defined in
equation~(\ref{eq:3}) follow also very well the same $L^{d}$
behavior, as expected.

In this subsection we have presented a reliable analysis of the
first-order transition features of the triangular Blume-Capel
model at the value $\Delta=2.95$ of the crystal field and we have
verified all the theoretical expectations of finite-size scaling
in first-order phase transitions. Additionally, we have estimated
important features of this transition, i.e. the surface tension
and latent heat of the transition, using the well-established
method of Lee and Kosterlitz~\cite{lee90}.

\subsection{Phase diagram and tricriticality}
\label{sec:3c}

\begin{figure}
\resizebox{1 \columnwidth}{!}{\includegraphics{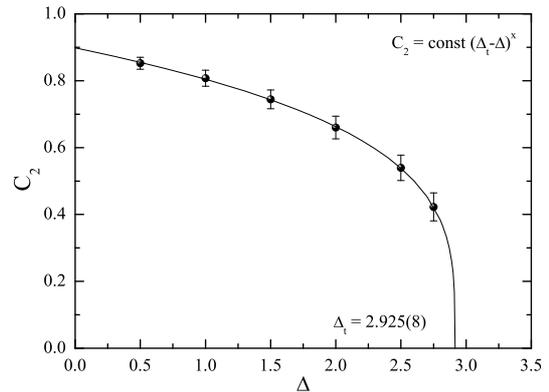}}
\caption{An approximate estimation of the tricritical value of the
crystal field by fitting the decreasing logarithmic amplitudes
$C_{2}$ of the model [see equation~(\ref{eq:5})] at suitable
values of the crystal field to a power law, as shown in the
figure.} \label{fig:13}
\end{figure}

We close this Section with our results on the phase diagram of the
model and the related issue of tricriticality. From
figure~\ref{fig:7} of subsection~\ref{sec:3a} one can observe the
expected Ising logarithmic divergence of the specific-heat maxima.
Avoiding the value $\Delta=0$, which suffers from (small)
crossover effects, we attempted to estimate the tricritical value
of the crystal field by fitting the decreasing logarithmic
amplitudes $C_{2}$ estimated in the simultaneous fitting of
figure~\ref{fig:7} to a suitable power law, as shown in the
figure~\ref{fig:13}. This may look like a questionable idea, since
the behavior of specific heat data is the Achilles' heel of
finite-size scaling analysis. Yet, figure~\ref{fig:13} shows that
besides the rather large errors in the logarithmic amplitudes
$C_{2}$, one may approximately estimate the tricritical
crystal-field value to be $\Delta_{t}=2.925(8)$, as shown in the
panel of the figure, in excellent agreement with the previous
analysis of the energy probability density functions and the
related thermodynamic quantities of figures~\ref{fig:1} and
\ref{fig:2}. We should note here that, such an idea was first
suggested and performed in the recent papers of
reference~\cite{malakis} for the cases of the pure and random-bond
square Blume-Capel model, where it has also produced quite
acceptable results. Furthermore, we aim here only to a qualitative
prediction of the tricritical crystal field value. A most accurate
prediction of these first-order transition features of the phase
diagram of the model should follow different routes,
well-established in the literature, which are, in any case, well
beyond the scope of the present paper.

\begin{figure}
\resizebox{1 \columnwidth}{!}{\includegraphics{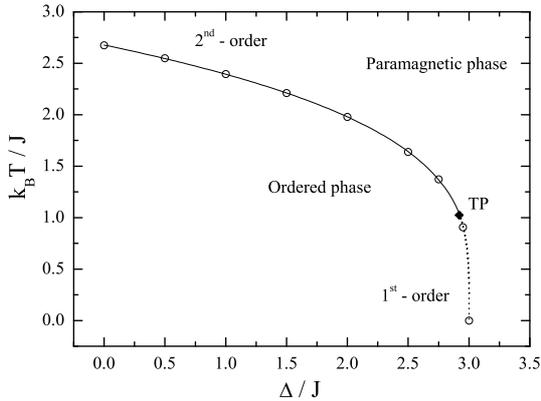}}
\caption{Approximation of the phase diagram of the triangular
Blume-Capel model. The dotted and solid curves correspond
respectively to first- and second-order phase transition lines
that separate the ordered and paramagnetic phases. The black
rhombus points the approximate location of the tricritical point
(TP).} \label{fig:14}
\end{figure}

Using now our estimates for the transition temperatures obtained
throughout this paper for all the values of the crystal field
considered we attempt to construct in figure~\ref{fig:14} an
approximation of the phase diagram of the model. We have fitted
the points shown in figure~\ref{fig:14} using the following
power-law ansatz
\begin{equation}
\label{eq:6} T \sim (\Delta_{0}-\Delta)^{x},
\end{equation}
where $\Delta_{0}$ in the above equation~(\ref{eq:6}) denotes the
crossing of the boundary to the horizontal axis ($T=0$) and should
take the value $3(=zJ/2)$ for the present case of the triangular
lattice ($z=6$). The solid and dotted curves that correspond to
second- and first-order phase transitions lines that separate the
ordered from the paramagnetic phase have been obtained from a
fitting of the form~(\ref{eq:6}) that gives the value $3.01(2)$
for $\Delta_{0}$, very close to the expected value, and a value
$x=0.274(9)$ for the exponent. Using now equation~(\ref{eq:6}) and
the estimate for the $\Delta_{t}$ from figure~\ref{fig:13} we
obtain an estimate for $T_{t}=1.025(10)$. Thus, the overall
estimate for the tricritical point is
$(\Delta_{t},T_{t})=(1.025(10),2.925(8))$ and is marked with a
black rhombus in figure~\ref{fig:14}. We should note here that,
although with the power-law fitting~(\ref{eq:6}) we get a very
nice estimate for the value $\Delta_{0}$ and also an excellent
concurrence between numerically estimated transition points and
the applied law, our attempt above aims only at a numerical
quantitative approximation for the main part of the diagram and
not at the correct asymptotic behavior at its ends. In fact, is
well known that the asymptotic approach of phase boundaries to the
sections with the axis follows specific power-law behaviors with
exponents related to the critical exponents describing the
transitions in these part of the diagram~\cite{aharony}.

\section{Conclusions}
\label{sec:4}

In the present paper we have performed a careful numerical
investigation of the critical properties of the Blume-Capel model
embedded in the triangular lattice. By applying an extensive
two-stage Wang-Landau entropic sampling we have studied the
finite-size scaling properties of the model for several values of
the crystal field in both its first- and second-order phase
transition regime and we have estimated with high accuracy
transition temperatures and critical exponents.

In particular, for the regime of second-order phase transitions we
have verified the theoretical expectation that the Blume-Capel
model belongs to the universality class of the simple $d=2$ Ising
model, whereas for the regime of first-order phase transitions we
have identified the most characteristic features signaling a
first-order phase transition and we have estimated the surface
tension and latent heat of the transition using the method of Lee
and Kosterlitz. Finally, using our estimates for the transition
temperatures and a novel scheme that takes advantage of the
logarithmic scaling behavior of the specific heat, we have
proposed an approximation of the phase diagram of the model in the
overall $(\Delta,T)$-plane and we have refined previous
mean-field-type estimates for the tricritical point.

Closing, we would like to leave as an open research challenge the
investigation of the phase diagram of the triangular Blume-Capel
model via a multi-parametric Wang-Landau method that would give
direct access to the estimation of the phase diagram. Such an
analysis would be complementary to the present contribution and
has been already successfully performed by Silva et
al.~\cite{silva} for the case of the square lattice Blume-Capel
model.

\begin{acknowledgement}
The author would like to thank Professor A.N. Berker for useful
discussions and a critical reading of the manuscript.
\end{acknowledgement}

\end{document}